# On a two-parameter extension of the lattice KdV system associated with an elliptic curve

*Frank W NIJHOFF* † *and Sian PUTTOCK* ‡

† *Department of Applied Mathematics, University of Leeds, Leeds LS2 9JT, United Kingdom*
*E-mail: frank@maths.leeds.ac.uk*
‡ *ibid.*
*E-mail: sian@maths.leeds.ac.uk*

**Abstract**

A general structure is developed from which a system of integrable partial difference equations is derived generalising the lattice KdV equation. The construction is based on an infinite matrix scheme with as key ingredient a (formal) elliptic Cauchy kernel. The consistency and integrability of the lattice system is discussed as well as special solutions and associated continuum equations.

## 1 Introduction

In recent years the integrability of discrete equations, i.e. ordinary or partial difference equations as well as analytic difference equations, has become an issue of considerable attention, (cf. the biannual sequel of SIDE meetings on *Symmetries and Integrability of Difference Equations*, cf. `http://www.maths.leeds.ac.uk/~side`). A particularly interesting development has been on the one hand the search for precise definitions of integrability in the discrete regime and the subsequent development of integrability detectors, on the other hand progress in the classification of various types of integrable discrete systems, such as the so-called discrete Painlevé equations.

One line of research since the mid-seventees with the pioneering works of Ablowitz and Ladik, [1] and of Hirota, [9], has been the study of integrable *lattice equations*, which are partial difference equations living on two-or higher dimensional space-time lattices. A systematic approach for the deriviation of such equations was performed in a series of papers, cf. e.g. [17, 23, 22] from the point of view of formal integral equations. The key ingredients in this approach, loosely coined *direct linearisation*, are two-fold: linear discrete or continuous dynamics residing in plane wave factors (the object $\rho(k)$ introduced below), and a formal Cauchy kernel ($\Omega(\kappa_1, \kappa_2)$ in the notation of the present paper) by which the connection with nonlinear objects takes place. The implementation of the method is in terms of an infnite matrix structure, which, albeit stricly formal, has proven to be powerful tool in setting up the basic relations from which all ingredients (Lax pairs, Bäcklund and Miura transformations, hierarchies of commuting flows, etc.) can be derived without having to rely on further *Ansätze*.

The resulting discrete systems on the 2D lattice are particularly important for developing an understanding of the notions of discrete integrability, in particular since other reduced systems (such as discrete Painlevé equations, integrable finite-dimensional mappings and discrete-time many-body systems) can be viewed as special solutions. Furthermore, there exist compelling connections with *difference geometry*, and the combinatorics on graphs.

The simplest example is possibly the lattice (potential) KdV equation

$$\left(a+b+u-\widehat{\widetilde{u}}\right)(a-b+\widehat{u}-\widetilde{u})=a^2-b^2 \ , \tag{1.1}$$

which, together with its companion equations and reductions, was studied at length in numerous papers, cf. e.g. [17, 22, 21, 7, 18].

The notation used to describe this lattice system is as follows: Denoting the dependent variable on the two-dimensional lattice by by $u=u_{n,m}$ $(n,m\in\mathbb{Z})$ the equation is given in terms of the variables $u$, $\widehat{u}$, $\widetilde{u}$ and $\widehat{\widetilde{u}}$ around an elementary plaquette as indicated in Figure 1.

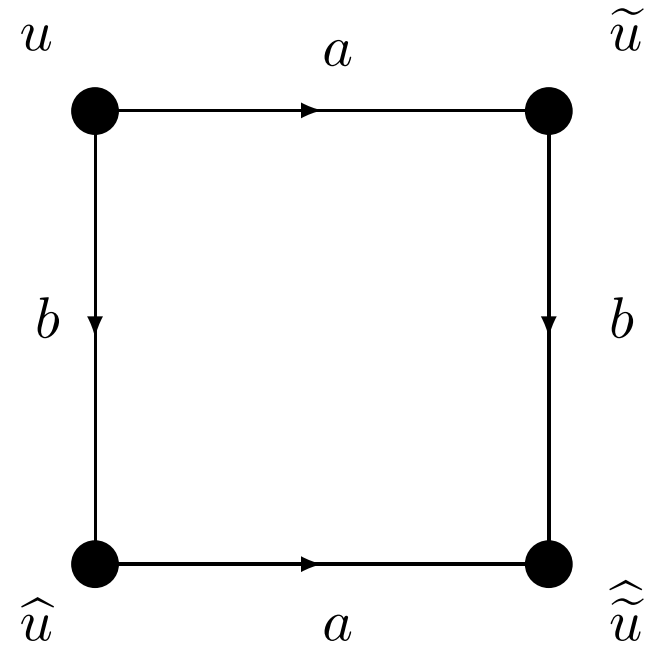

**Figure 1:** *Elementary quadrilateral on which the lattice equation is defined.*

The integrability of such lattice equations can be understood in a rather simple way: it seems entirely to reside in a simple but deep combinatorial property, first described in the paper [19], namely that two-dimensional lattice equations in fact form parameter-families of compatible equations which can be consistently embedded in a multidimensional lattice, on each two-dimensional sublattice of which a copy of the lattice equation can be defined. As was shown in [19, 13], cf. also [4], this property is powerful enough to derive subsequently Lax pairs for the lattice equations, which can then be used to study the analytic properties of solutions. More recently in [3] the property was used to arrive at a full classification of lattice equations integrable in this sense under certain some additional simplifying assumptions. In particular, the most parameter-rich equation emerging from this classification was a lattice system already found by V. Adler in [2], for which the Lax pair was derived in [13].

Although these developments form a considerable step forward, relatively little is known of the underlying algebraic and analytic structures of Adler's lattice equations (this being the subject of a paper [15] in preparation). In this note we present another integrable lattice system which is naturally associated with an elliptic curve, and which is given by the following coupled system of equations:

$$\left(a+b+u-\widehat{\widetilde{u}}\right)(a-b+\widetilde{u}-\widehat{u}) = a^2-b^2+f\,(\widetilde{s}-\widehat{s})\left(\widehat{\widetilde{s}}-s\right) \tag{1.2a}$$

$$\left(\widehat{\widetilde{s}}-s\right)(\widetilde{w}-\widehat{w}) = [(a+u)\widetilde{s}-(b+u)\widehat{s}]\,\widehat{\widetilde{s}}-\left[(a-\widehat{\widetilde{u}})\widehat{s}-(b-\widehat{\widetilde{u}})\widetilde{s}\right]s \tag{1.2b}$$

$$(\widehat{s}-\widetilde{s})\left(\widehat{\widetilde{w}}-w\right) = \left[(a-\widetilde{u})s+(b+\widetilde{u})\widehat{\widetilde{s}}\right]\widehat{s}-\left[(a+\widehat{u})\widehat{\widetilde{s}}+(b-\widehat{u})s\right]\widetilde{s} \tag{1.2c}$$

$$\left(a+u-\frac{\widetilde{w}}{\widetilde{s}}\right)\left(a-\widetilde{u}+\frac{w}{s}\right) = a^2-P(s\widetilde{s}) \tag{1.2d}$$

$$\left(b+u-\frac{\widehat{w}}{\widehat{s}}\right)\left(b-\widehat{u}+\frac{w}{s}\right) = b^2-P(s\widehat{s}) \tag{1.2e}$$

in which

$$P(x) \equiv \frac{1}{x}+3e+fx\ ,$$

with $e$ and $f$ moduli of an elliptic curve $y^2 = P(x)$ . We will show that this system (1.2) is a natural extension of the lattice KdV equation (1.1), the latter being obviously recovered when $f=0$, i.e. when the elliptic curve degenerates.

The system (1.2) is admittedly rather complicated, and is outside the class of systems considered in the paper [3]. Rather than starting with an ad-hoc *Ansatz* we want to derive the system in a systematic way from a structure that not only provides us with the system but also allows us to construct its Lax pair and its linearisation. In order to achieve this we start from an infinite matrix system with a quasigradation of elliptic type similar to the ones that occur in the Krichever-Novikov algebras. The nonlinear structure arises from objects in this algebra defined through an elliptic Cauchy kernel. As a result we find a system of equations which forms a two-parameter "elliptic" deformation of the lattice systems of KdV type investigated in the past, cf. e.g. [17, 23, 14]. The extended lattice systems then will serve as a starting point for the study of a number of new discrete systems, notably the ones that arise from finite-dimensional or similarity reductions. Also their continuum counterparts form an interesting new class of integrable systems that merit further investigations.

# 2 Infinite Matrix Structure

## 2.1 Elliptic Matrices

The starting point for the construction of the elliptic discrete model is an algebra $\mathcal{A} = \mathrm{Mat}^0_{\mathbb{Z}}(\mathbb{C})$ of centred infinite "elliptic" matrices. We prefer to describe this algebra in simple terms without going into details of the construction, as we are only interested in some basic relations.

We mean by $\mathcal{A}$ the associative algebra (with unit $\mathbf{1}$) with the usual matrix multiplication relative to a designated central entry, i.e. elements $\boldsymbol{A},\boldsymbol{B}\in\mathcal{A}$ are infinite matrices $\boldsymbol{A} = (A_{i,j})$, $\boldsymbol{B} = (B_{i,j})$, $(i,j\in\mathbb{Z})$, having $A_{0,0}$, respectively $B_{0,0}$ as central entry with respect to which the matrix multiplication is defined, i.e. through the usual formula

$$(\boldsymbol{A}\cdot\boldsymbol{B})_{ij}=\sum_{k\in\mathbb{Z}}A_{i,k}B_{k,j}$$

Since we will only be interested in the formal structure leading eventually to nonlinear discrete equations, we will not be distracted by questions related to the convergence of infinite sums (in which case one has to impose some additional restrictions on the matrices, which, however, will not be relevant for the present discussion). The main property of this elliptic matrix algebra is that it is not graded in the usual way (i.e. through the organisation of rows and columns according to integer steps), but quasigraded by means of two types of index raising operators $\boldsymbol{\Lambda}$ (of order 1) and $\boldsymbol{P}$ (of order 2), subject to the algebraic relation

$$\boldsymbol{\Lambda}^2=\boldsymbol{P}+3e\mathbf{1}+f\boldsymbol{P}^{-1}\qquad,\qquad \boldsymbol{P}\cdot\boldsymbol{\Lambda}=\boldsymbol{\Lambda}\cdot\boldsymbol{P}\ , \tag{2.1}$$

where $e,f\in\mathbb{C}$ are moduli of an elliptic curve given by the relation (2.1). This means that the indices of the matrices are defined relative to the action of these operators, namely as follows: for each element $\boldsymbol{A}=(A_{ij})\in\mathcal{A}$ we identify the indices as follows:

$$\begin{aligned}
A_{2i,2j}&=\left(\boldsymbol{P}^i\cdot\boldsymbol{A}\cdot{}^t\boldsymbol{P}^j\right)_{00}\\
A_{2i+1,2j}&=\left(\boldsymbol{P}^i\cdot\boldsymbol{\Lambda}\cdot\boldsymbol{A}\cdot{}^t\boldsymbol{P}^j\right)_{00}\\
A_{2i,2j+1}&=\left(\boldsymbol{P}^i\cdot\boldsymbol{A}\cdot{}^t\boldsymbol{\Lambda}\cdot{}^t\boldsymbol{P}^j\right)_{00}\\
A_{2i+1,2j+1}&=\left(\boldsymbol{P}^i\cdot\boldsymbol{\Lambda}\cdot\boldsymbol{A}\cdot{}^t\boldsymbol{\Lambda}\cdot{}^t\boldsymbol{P}^j\right)_{00}
\end{aligned}\tag{2.2}$$

We need three ingredients to formulate the scheme:

- A parameter-family of elements $\boldsymbol{C}\in\mathcal{A}$ which encodes the dynamics of the system via the linear relations

$$\widetilde{\boldsymbol{C}}\cdot\left(a-{}^t\boldsymbol{\Lambda}\right)\ =\ (a+\boldsymbol{\Lambda})\cdot\boldsymbol{C}\ , \tag{2.3a}$$
$$\widehat{\boldsymbol{C}}\cdot\left(b-{}^t\boldsymbol{\Lambda}\right)\ =\ (b+\boldsymbol{\Lambda})\cdot\boldsymbol{C}\ . \tag{2.3b}$$

  In addition we can impose differential relations with respect to a hierarchy of continuous flow-variables $x_j$:

$$\frac{\partial}{\partial x_j}\boldsymbol{C}=\boldsymbol{\Lambda}^j\cdot\boldsymbol{C}-\boldsymbol{C}(-{}^t\boldsymbol{\Lambda})^j\quad,\quad j\in\mathbb{Z}\ . \tag{2.4}$$

- The formal elliptic Cauchy kernel $\boldsymbol{\Omega}\in\mathcal{A}$ which obeys the following defining equations:

$$\boldsymbol{\Omega}\cdot\boldsymbol{\Lambda}+{}^t\boldsymbol{\Lambda}\cdot\boldsymbol{\Omega}\ =\ \boldsymbol{O}-f\,{}^t\boldsymbol{P}^{-1}\cdot\boldsymbol{O}\cdot\boldsymbol{P}^{-1}\ , \tag{2.5a}$$
$$\boldsymbol{\Omega}\cdot\boldsymbol{P}-{}^t\boldsymbol{P}\cdot\boldsymbol{\Omega}\ =\ \boldsymbol{O}\cdot\boldsymbol{\Lambda}-{}^t\boldsymbol{\Lambda}\cdot\boldsymbol{O}\ . \tag{2.5b}$$

  in which $\boldsymbol{O}$ is the projection matrix on the central element, i.e. $(\boldsymbol{O}\cdot\boldsymbol{A})_{i,j}=\delta_{i,0}\boldsymbol{A}_{0,j}$ and $(\boldsymbol{A}\cdot\boldsymbol{O})_{i,j}=\delta_{0,j}\boldsymbol{A}_{i,0}$.

## 2.2 Symbols of the operators

The infinite matrix structure described above is most easily explained in terms of the *symbols* of the operators involved. These arise when we consider the action of these operators on a basis of monomials, i.e. on an infinite vector $\boldsymbol{c}(\kappa)$ with entries $(\boldsymbol{c}(\kappa))_j = \kappa^j$ . If $\Lambda$ denotes a raising operator acting naturally on these vectors by $\Lambda \cdot \boldsymbol{c}(\kappa) = \kappa \boldsymbol{c}(\kappa)$ and if we allow for the formal expressions $\boldsymbol{P} = \wp(\Lambda) - e\mathbf{1}$ , $2\boldsymbol{\Lambda} \cdot \boldsymbol{P} = \wp'(\Lambda)$ , with $\wp(\kappa) = \wp(\kappa|2\omega, 2\omega')$ being the Weierstrass $\wp$-function with periods $2\omega$,$2\omega'$, and where $e = \wp(\omega)$ . In this way we obtain a realisation of the operators $\boldsymbol{\Lambda}$ and $\boldsymbol{P}$ acting on the vector $\boldsymbol{c}(\kappa)$ as follows:

$$\boldsymbol{\Lambda} \cdot \boldsymbol{c}(\kappa) = k\boldsymbol{c}(\kappa) \qquad , \qquad \boldsymbol{P} \cdot \boldsymbol{c}(\kappa) = K\boldsymbol{c}(\kappa) \ ,$$

thus leading to a realisation of the left- index-raising operators:

$$\boldsymbol{\Lambda} \quad \leftrightarrow \quad \zeta(\kappa_1 + \omega) - \zeta(\kappa_1) - \zeta(\omega) = \frac{1}{2}\frac{\wp'(\kappa_1)}{\wp(\kappa_1) - e} \qquad , \qquad \boldsymbol{P} \quad \leftrightarrow \quad \wp(\kappa_1) - e$$

(where $\kappa_1$ and $\kappa_2$ can be thought of as the symbols for the matrices $\Lambda$ resp. ${}^t\Lambda$ corresponding to the rational case). Similarly we can realise their transposeds by their action on a corresponding transposed vector ${}^t\boldsymbol{c}(\kappa)$, leading to

$${}^t\boldsymbol{\Lambda} \quad \leftrightarrow \quad \zeta(\kappa_2 + \omega) - \zeta(\kappa_2) - \zeta(\omega) = \frac{1}{2}\frac{\wp'(\kappa_2)}{\wp(\kappa_2) - e} \qquad , \qquad {}^t\boldsymbol{P} \quad \leftrightarrow \quad \wp(\kappa_2) - e \ .$$

In the above $\sigma$, $\zeta$ are the well-known Weierstrass functions, and where $\omega$, $\omega'$ are the corresponding *half-periods*. Setting $\omega'' = -\omega - \omega'$, then $e = \wp(\omega)$, $e' = \wp(\omega')$, $e'' = \wp(\omega'')$ are the corresponding branch points of the Weierstrass curve:

$$\mu^2 = 4(\lambda - e)(\lambda - e')(\lambda - e'') \ .$$

For our later purpose we prefer to write the curve in the following form:

$$k^2 = K + 3e + \frac{f}{K} \tag{2.6}$$

in terms of cordinates $(k, K)$ paramtrised by $K = \wp(\kappa) - e$ , $2kK = \wp'(\kappa)$ , and where $f = (e - e')(e - e'')$ . In this way we arrive at the symbols of these operators in terms of which we have in particular the realisation of the Caucy kernel:

$$\boldsymbol{\Omega} \quad \leftrightarrow \quad \Omega(\kappa_1, \kappa_2) = \frac{k_1 - k_2}{K_1 - K_2} = \frac{1 - f/(K_1K_2)}{k_1 + k_2} \tag{2.7}$$

with the identifications:

$$k_i = \frac{1}{2}\frac{\wp'(\kappa_i)}{\wp(\kappa_i) - e} \quad , \quad K_i = \wp(\kappa_i) - e \quad (i = 1, 2) \ . \tag{2.8}$$

In order to etsablish some of the results for the Cauchy kernel we need some of the well-known addition formulae for the Weierstrass functions. The main identities that we need in the context of the present paper are the following ones:

$$(\zeta(\kappa + \omega) - \zeta(\kappa) - \zeta(\omega))^2 = \wp(\kappa + \omega) + \wp(\kappa) + e \ ,$$

$$(\wp(\kappa + \omega) - e)\,(\wp(\kappa) - e) = (e - e')(e - e'') = f \ .$$

### 2.3 Basic relations

The main object of interest, i.e. the one for which we obtain nonlinear equations, is the following infinite matrix $\boldsymbol{U} \in \mathcal{A}$:

$$\boldsymbol{U} \equiv \boldsymbol{C} \cdot (\mathbf{1} + \boldsymbol{\Omega} \cdot \boldsymbol{C})^{-1} \ , \tag{2.9}$$

the components of which are denoted by $U_{i,j}$, $(i, j \in \mathbb{Z})$.

In order to obtain the relevant Lax pairs, we introduce also the infinite-component vectors $\boldsymbol{u}(\kappa)$

$$\boldsymbol{u}(\kappa) = (\mathbf{1} - \boldsymbol{U} \cdot \boldsymbol{\Omega}) \cdot \boldsymbol{c}(\kappa)\rho(k) \ , \tag{2.10}$$

with components $(\boldsymbol{u}(\kappa))_j$, $(j \in \mathbb{Z})$, in which $\rho(k)$ is a scalar function of discrete variables $n, m \in \mathbb{Z}$ as follows

$$\rho(k) = \rho_{n,m}(k) = \left(\frac{a+k}{a-k}\right)^n \left(\frac{b+k}{b-k}\right)^m \rho_{0,0}(k) \ , \tag{2.11}$$

Let us now present the basic equations resulting from this scheme. For the infinite matrix $\boldsymbol{U}$ we can derive the discrete matrix Riccati type of relations

$$\widetilde{\boldsymbol{U}} \cdot (a - {}^t\boldsymbol{\Lambda}) = (a + \boldsymbol{\Lambda}) \cdot \boldsymbol{U} - \widetilde{\boldsymbol{U}} \cdot \left(\boldsymbol{O} - f\,{}^t\boldsymbol{P}^{-1} \cdot \boldsymbol{O} \cdot \boldsymbol{P}^{-1}\right) \cdot \boldsymbol{U} \ , \tag{2.12a}$$

$$\widehat{\boldsymbol{U}} \cdot (b - {}^t\boldsymbol{\Lambda}) = (b + \boldsymbol{\Lambda}) \cdot \boldsymbol{U} - \widehat{\boldsymbol{U}} \cdot \left(\boldsymbol{O} - f\,{}^t\boldsymbol{P}^{-1} \cdot \boldsymbol{O} \cdot \boldsymbol{P}^{-1}\right) \cdot \boldsymbol{U} \ , \tag{2.12b}$$

together with the algebraic relations

$$\boldsymbol{U} \cdot {}^t\boldsymbol{P} = \boldsymbol{P} \cdot \boldsymbol{U} - \boldsymbol{U} \cdot \left(\boldsymbol{O} \cdot \boldsymbol{\Lambda} - {}^t\boldsymbol{\Lambda} \cdot \boldsymbol{O}\right) \cdot \boldsymbol{U} \qquad \Leftrightarrow \tag{2.13a}$$

$$\boldsymbol{U} \cdot {}^t\boldsymbol{P}^{-1} = \boldsymbol{P}^{-1} \cdot \boldsymbol{U} + \boldsymbol{U} \cdot {}^t\boldsymbol{P}^{-1} \cdot \left(\boldsymbol{O} \cdot \boldsymbol{\Lambda} - {}^t\boldsymbol{\Lambda} \cdot \boldsymbol{O}\right) \cdot \boldsymbol{P}^{-1} \cdot \boldsymbol{U} \ , \tag{2.13b}$$

the latter two equations (which do not involve lattice shifts) being needed for the sake of consistency.

In addition for the continuous case the set of continuous equations can be derived if one includes a factor $\exp(2\sum_{j \text{ odd}} k^j x_j)$ on the r.h.s. of (2.11), yielding:

$$\frac{\partial}{\partial x_j}\boldsymbol{U} = \boldsymbol{\Lambda}^j \cdot \boldsymbol{U} + \boldsymbol{U} \cdot {}^t\boldsymbol{\Lambda}^j - \boldsymbol{U} \cdot \left(\boldsymbol{O}_j - f\,{}^t\boldsymbol{P}^{-1} \cdot \boldsymbol{O}_j \cdot \boldsymbol{P}^{-1}\right) \cdot \boldsymbol{U} \quad , \quad (j \ \text{ odd}) \ , \tag{2.14}$$

in which

$$\boldsymbol{O}_j = \sum_{i=0}^{j-1} (-\,{}^t\boldsymbol{\Lambda})^i \cdot \boldsymbol{O} \cdot \boldsymbol{\Lambda}^{j-1-i} \ .$$

These equations can be shown to compatible with the discrete relations (2.12) and lead by themselves to a hierarchy of commuting continuous flows.

In the next section we will derive from the infinite matrix relations (2.12) and (2.14), together with (2.13), *closed-form* partial difference equations (in terms of the lattice shifts $\widetilde{\cdot}$ and $\widehat{\cdot}$) or partial differential equations (in terms of the derivatives with respect to $a$ and $b$) or mixed differential-difference equations, but in terms of specific well-chosen entries of the infinite matrix $\boldsymbol{U}$. Having obtained these equations by means of the infinite matrix structure, subsequently the latter formal tool can be set aside, and the integrability can be verified on its own merit. However, in order to obtain additional structures, such as for instance the Lax pairs, the infinite matrix structure allows to construct these as well in a systematic way.

Thus, from (2.10) and the form of $\rho(k)$, by using the equations (2.12) and (2.13) together with the properties (2.5a) of the infinite matrix kernel $\mathbf{\Omega}$ one can derive the basic set of equations for the vector $\boldsymbol{u}(\kappa)$. These constitute the *linear* discrete relations

$$(a-k)\widetilde{\boldsymbol{u}}(\kappa) = \Big[(a+\mathbf{\Lambda}) - \widetilde{\boldsymbol{U}}\cdot\big(\boldsymbol{O} - f\,{}^t\boldsymbol{P}^{-1}\cdot\boldsymbol{O}\cdot\boldsymbol{P}^{-1}\big)\Big]\cdot\boldsymbol{u}(\kappa)\ , \tag{2.15a}$$

$$(b-k)\widehat{\boldsymbol{u}}(\kappa) = \Big[(b+\mathbf{\Lambda}) - \widehat{\boldsymbol{U}}\cdot\big(\boldsymbol{O} - f\,{}^t\boldsymbol{P}^{-1}\cdot\boldsymbol{O}\cdot\boldsymbol{P}^{-1}\big)\Big]\cdot\boldsymbol{u}(\kappa)\ . \tag{2.15b}$$

as well as the linear "algebraic" relations

$$K\,\boldsymbol{u}(\kappa) = \big[\boldsymbol{P} - \boldsymbol{U}\cdot\big(\boldsymbol{O}\cdot\mathbf{\Lambda} - {}^t\mathbf{\Lambda}\cdot\boldsymbol{O}\big)\big]\cdot\boldsymbol{u}(\kappa) \qquad \Leftrightarrow \tag{2.16a}$$

$$\frac{1}{K}\,\boldsymbol{u}(\kappa) = \big[\boldsymbol{P}^{-1} + \boldsymbol{U}\cdot{}^t\boldsymbol{P}^{-1}\cdot\big(\boldsymbol{O}\cdot\mathbf{\Lambda} - {}^t\mathbf{\Lambda}\cdot\boldsymbol{O}\big)\cdot\boldsymbol{P}^{-1}\big]\cdot\boldsymbol{u}(\kappa)\ , \tag{2.16b}$$

where $K$ and $k$ are related through the elliptic curve $k^2 = P(K)$ . Similarly, we have linear differential relations for $\boldsymbol{u}(\kappa)$ as a consequence of (2.14), namely

$$\frac{\partial}{\partial x_j}\boldsymbol{u}(\kappa) = \mathbf{\Lambda}^j\cdot\boldsymbol{u}(\kappa) + k^j\,\boldsymbol{u}(\kappa) - \boldsymbol{U}\cdot\big(\boldsymbol{O}_j - f\,{}^t\boldsymbol{P}^{-1}\cdot\boldsymbol{O}_j\cdot\boldsymbol{P}^{-1}\big)\cdot\boldsymbol{u}(\kappa) \quad , \quad (j\ \text{odd})\ . \tag{2.17}$$

From these equations one can eventually derive all the relevant Lax pairs, both for the continuum equations as well as on the lattice.

**Remark:** The KdV class of systems by the additional requirement that the matrix $\boldsymbol{C}$ and hence $\boldsymbol{U}$ is *symmetric* under transposition, i.e.

$${}^t\boldsymbol{C} = \boldsymbol{C} \qquad \Rightarrow \qquad {}^t\boldsymbol{U} = \boldsymbol{U}$$

which is equivalent to saying that its entries obey: $U_{i,j} = U_{j,i}$. This will be used in the next section in the selection of the components that will enter the main equations.

## 3 Elliptic Lattice System

Having obtained the basic equations (2.12) and (2.14), together with (2.13), in terms of the infinite matrix $\boldsymbol{U}$ we now derive closed-form equations in terms of its entries. It turns out that the following choice of entries leads to closed-form equations by combining the relations associated with two different lattice shifts, namely

$$u = U_{0,0} \quad , \quad s = U_{-2,0} \quad , \quad h = U_{-2,-2}$$
$$v = 1 - U_{-1,0} \quad , \quad w = 1 + U_{-2,1} \quad ,$$

recalling that $U_{i,j} = U_{j,i}$. Using the identification of the entries (2.1) and the relation (2.1) between the index-raising operators $\mathbf{\Lambda}$ and $\boldsymbol{P}$ (and similar relation for their transposed), we can derive the following set of relations:

$$a - fh = \frac{(a-\widetilde{u})s - \widetilde{v} + w}{\widetilde{s}} \tag{3.1a}$$

$$
\begin{aligned}
a + f\widetilde{h} &= \frac{(a+u)\widetilde{s} + v - \widetilde{w}}{s} &\quad (3.1b)\\
U_{1,-1} &= \frac{1 - vw}{s} &\quad (3.1c)\\
\widetilde{U}_{-1,-2} + U_{-1,-2} &= a(\widetilde{\widetilde{h}} - h) - f\widetilde{h}h + \widetilde{s}s &\quad (3.1d)\\
a(v - \widetilde{v}) &= fvh + fs\left(U_{-2,-1} + \widetilde{U}_{-2,-1}\right) + \widetilde{U}_{-1,1} + 3es + \widetilde{v}u &\\
&= f\widetilde{v}\widetilde{\widetilde{h}} + f\widetilde{s}\left(U_{-2,-1} + \widetilde{U}_{-2,-1}\right) + U_{-1,1} + 3e\widetilde{s} + v\widetilde{u} &\quad (3.1e)\\
a(\widetilde{w} - w) &= f\widetilde{h}w - \widetilde{s}\left(U_{0,1} + \widetilde{U}_{0,1}\right) + U_{-1,1} + 3e\widetilde{s} + \widetilde{w}\widetilde{u} &\\
&= fh\widetilde{w} - s\left(U_{0,1} + \widetilde{U}_{0,1}\right) + \widetilde{U}_{-1,1} + 3es + wu &\quad (3.1f)\\
U_{0,1} + \widetilde{U}_{0,1} &= a(\widetilde{u} - u) + \widetilde{u}u - f\widetilde{s}s &\quad (3.1g)
\end{aligned}
$$

and similar relations with $a$ replaced by $b$ and $\widetilde{\cdot}$ replaced by $\widehat{\cdot}$.

Combining the various relations we can extract from them the closed-form system of partial difference equations (1.2) which involves only the variables $u$, $s$ and $w$. Alternatively, we can derive also a lattice system in terms of $h$, $s$ and $v$, which reads

$$\left(a + b + f\widehat{\widetilde{h}} - fh\right)\left(a - b + f\widetilde{h} - f\widehat{h}\right) = a^2 - b^2 + f\left(\widetilde{s} - \widehat{s}\right)\left(\widehat{\widetilde{s}} - s\right) \quad (3.2a)$$

$$\left(\widehat{\widetilde{s}} - s\right)\left(\widehat{v} - \widetilde{v}\right) = \left[(a - fh)\widetilde{s} - (b - fh)\widehat{s}\right]\widehat{\widetilde{s}} - \left[(a + f\widehat{\widetilde{h}})\widehat{s} - (b + f\widehat{\widetilde{h}})\widetilde{s}\right] s \quad (3.2b)$$

$$\left(\widetilde{s} - \widehat{s}\right)\left(\widehat{\widetilde{v}} - v\right) = \left[(a + f\widetilde{h})s + (b - f\widetilde{h})\widehat{\widetilde{s}}\right]\widehat{s} - \left[(a - f\widehat{h})\widehat{\widetilde{s}} + (b + f\widehat{h})s\right]\widetilde{s} \quad (3.2c)$$

$$\left(a - fh + \frac{\widetilde{v}}{\widetilde{s}}\right)\left(a + f\widetilde{h} - \frac{v}{s}\right) = a^2 - P(s\widetilde{s}) \quad (3.2d)$$

$$\left(b - fh + \frac{\widehat{v}}{\widehat{s}}\right)\left(b + f\widehat{h} - \frac{v}{s}\right) = b^2 - P(s\widehat{s}) \;, \quad (3.2e)$$

and which is obviously equivalent to the lattice system (1.2).

If the parameter $f = 0$ the curve is degenerate and the first equation (1.2a) will decouple, yielding the lattice potential KdV equation (1.1) which was at the centre of previous derivations, cf. e.g. [17]-[14]. Thus, the system (1.2) can be viewed as a 2-parameter deformation of the latter equation. In what follows we demonstrate the integrability of the elliptic lattice system (1.2), and discuss that the systems is well-posed from the point of view of initial value problems. Obviously, we expect that all known results on finite-dimensional and similarity reductions that were performed for the lattice potential KdV equation can be extended in a natural way to the elliptic lattice system.

## 3.1 Lax pairs

From the relations (2.15) together with (2.16) one can *derive* now a Lax pair for the lattice system (1.2). In fact, by taking

$$\phi = \begin{pmatrix} (\boldsymbol{u}(\kappa))_0 \\ (\boldsymbol{u}(\kappa))_1 \end{pmatrix} \;, \quad (3.3)$$

the following discrete linear systems is obtained:

$$(a-k)\widetilde{\phi} = L(K)\phi \tag{3.4a}$$
$$(b-k)\widehat{\phi} = M(K)\phi \tag{3.4b}$$

with the points $(k,K)$ on the elliptic curve representing the spectral parameter. In (3.4) the matrices $L$ and $M$ are given by:

$$L(K) = \begin{pmatrix} a-\widetilde{u}+\frac{f}{K}\widetilde{s}w & 1-\frac{f}{K}\widetilde{s}s \\ K+3e-a^2+f\widetilde{s}s & \\ +(a-\widetilde{u})(a+u)+\frac{f}{K}\widetilde{w}w & a+u-\frac{f}{K}\widetilde{w}s \end{pmatrix} \tag{3.5}$$

and for $M$ a similar expression obtained from (3.5) by replacing $a$ by $b$ and $\widehat{\cdot}$ by $\widetilde{\cdot}$. It is straightforward to show that the discrete Lax equation arising from the compatibility condition of the linear system (3.4)

$$\widetilde{L}M = \widehat{M}L$$

gives rise to the set of equations (1.2). We observe that the matrices $L$ and $M$ depend rationally on $K$ only, and thus we do not seem to have essentially a spectral variable on the torus. Nevertheless, the solutions seem to depend essentially on the elliptic curve as we shall show below by means of soliton type solutions[1].

The Lax pair form the dual system (3.2) is derived in a similar way taking the vector

$$\psi = \begin{pmatrix} (\boldsymbol{u}(\kappa))_{-1} \\ (\boldsymbol{u}(\kappa))_{-2} \end{pmatrix} , \tag{3.6}$$

for which we have a similar linear system:

$$(a-k)\widetilde{\psi} = \mathcal{L}(K)\psi \tag{3.7a}$$
$$(b-k)\widehat{\psi} = \mathcal{M}(K)\psi \tag{3.7b}$$

where

$$\mathcal{L}(K) = \begin{pmatrix} a-fh+K\widetilde{v}s & \frac{f}{K}+3e-a^2+f\widetilde{s}s \\ & +(a+f\widetilde{h})(a-fh)+K\widetilde{v}v \\ 1-K\widetilde{s}s & a+f\widetilde{h}-K\widetilde{s}v \end{pmatrix} \tag{3.8}$$

and again for $\mathcal{M}$ a similar expression obtained from (3.8) by making the obvious replacements. The two Lax pairs (3.4) and (3.7) are gauge-related via the gauge transformation:

$$\phi = K\boldsymbol{G}\psi \quad , \quad \boldsymbol{G} = \begin{pmatrix} s & v \\ w & \frac{vw-1}{s} \end{pmatrix} \tag{3.9}$$

In fact, the two systems (1.2) and (3.2) are essentially the same system, albeit in terms of different variables. We note that the gauge conditions on the Lax matrices, e.g.

$$\widetilde{\boldsymbol{G}}\mathcal{L} = L\boldsymbol{G} , \tag{3.10}$$

leads to the relations in the (3.1).

[1]This seems to be in line with a recent observation by Bordag and Yanovski in [5] where it was shown that even for the Landau-Lifschitz equations one can have a Lax pair with polynomial dependence on the spectral variable.

### 3.2 Soliton type solutions

It is relatively straightforward from the infinite matrix structure to construct soliton solutions, using the symbol representation presented in the previous section.

Introducing the $N \times N$ matrix $\boldsymbol{M}$ with entries

$$M_{ij} = \frac{1 - f/(K_i K_j)}{k_i + k_j} r_i \quad , \quad (i,j = 1, \dots, N) \tag{3.11}$$

where the parameters of the solution $(k_i, K_i)$ are points on the elliptic curve:

$$k^2 = K + 3e + \frac{f}{K} \ ,$$

and the vector $\boldsymbol{r} = (r_i)_{i=1,\dots,N}$ with components

$$r_i = \left(\frac{a + k_i}{a - k_i}\right)^n \left(\frac{b + k_i}{b - k_i}\right)^m r_i^0 \ , \tag{3.12}$$

where the coefficients $r_i^0$ are independent of $n$, $m$. In this case we can take for the the infinite matrix $\boldsymbol{C}$ a finite-rank matrix of the form:

$$\boldsymbol{C} = \sum_{i=1}^{N} r_i \boldsymbol{c}_{\kappa_i} {}^t\boldsymbol{c}_{\kappa_i} \ , \tag{3.13}$$

and this leads to the following explicit formulae for the quantities of interest:

$$u = \boldsymbol{e} \cdot (\boldsymbol{1} + \boldsymbol{M})^{-1} \cdot \boldsymbol{r} \tag{3.14a}$$
$$s = \boldsymbol{e} \cdot \boldsymbol{K}^{-1} \cdot (\boldsymbol{1} + \boldsymbol{M})^{-1} \cdot \boldsymbol{r} \tag{3.14b}$$
$$w = 1 + \boldsymbol{e} \cdot \boldsymbol{K}^{-1} \cdot (\boldsymbol{1} + \boldsymbol{M})^{-1} \cdot \boldsymbol{k} \cdot \boldsymbol{r} \tag{3.14c}$$
$$v = 1 - \boldsymbol{e} \cdot \boldsymbol{K}^{-1} \cdot \boldsymbol{k} \cdot (\boldsymbol{1} + \boldsymbol{M})^{-1} \cdot \boldsymbol{r} \tag{3.14d}$$
$$h = \boldsymbol{e} \cdot \boldsymbol{K}^{-1} \cdot (\boldsymbol{1} + \boldsymbol{M})^{-1} \cdot \boldsymbol{K}^{-1} \cdot \boldsymbol{r} \tag{3.14e}$$

in which have employed the vector $\boldsymbol{e} = (1, 1, \dots, 1)$ and the diagonal matrices

$$\boldsymbol{K} = \mathrm{diag}(K_1, K_2, \dots, K_N) \quad , \quad \boldsymbol{k} = \mathrm{diag}(k_1, k_2, \dots, k_N) \ .$$

We note that although the dynamics itself (encoded in the wave factors $r_i$) does not involve the elliptic curve, the soliton solutions essentially depend on the variables on the curve. In fact, it is easily verified by direct calculation that the formulae (3.14) provide a solution to the lattice system (1.2) if and only if the elliptic curve relation holds between the parameters $k_i$ and the parameters $K_i$.

## 4 Initial value problems on the lattice

We now address the question of how to define a well-posed initial value problem (IVP) for the lattice systems (1.2) and, equivalently, for the dual lattice system (3.2).

Rewriting the lattice system (1.2) as follows:

$$\left(a + b + u - \widehat{\widetilde{u}}\right)(a - b + \widehat{u} - \widetilde{u}) = a^2 - b^2 + f\,(\widetilde{s} - \widehat{s})\left(\widehat{\widetilde{s}} - s\right) \tag{4.1a}$$

$$\left[\left(a+u-\frac{\widetilde{w}}{\widetilde{\widetilde{s}}}\right)\widetilde{s}-\left(b+u-\frac{\widehat{w}}{\widehat{\widehat{s}}}\right)\widehat{s}\right]\widehat{\widetilde{s}}=\left[\left(a-\widehat{\widetilde{u}}+\frac{\widehat{w}}{\widehat{s}}\right)\widehat{s}-\left(b-\widehat{\widetilde{u}}+\frac{\widetilde{w}}{\widetilde{s}}\right)\widetilde{s}\right]s \tag{4.1b}$$

$$\left[\left(a-\widetilde{u}+\frac{w}{s}\right)s+\left(b+\widetilde{u}-\frac{\widehat{\widetilde{w}}}{\widehat{\widetilde{s}}}\right)\widehat{\widetilde{s}}\right]\widehat{s}=\left[\left(a+\widehat{u}-\frac{\widehat{\widetilde{w}}}{\widehat{\widetilde{s}}}\right)\widehat{\widetilde{s}}+\left(b-\widehat{u}+\frac{w}{s}\right)s\right]\widetilde{s} \tag{4.1c}$$

$$\left(a+u-\frac{\widetilde{w}}{\widetilde{\widetilde{s}}}\right)\left(a-\widetilde{u}+\frac{w}{s}\right)=a^2-P(s\widetilde{s}) \tag{4.1d}$$

$$\left(b+u-\frac{\widehat{w}}{\widehat{\widehat{s}}}\right)\left(b-\widehat{u}+\frac{w}{s}\right)=b^2-P(s\widehat{s}) \tag{4.1e}$$

where as before $P(x) = 1/x + 3e + fx$ , we have to assign values to the dependent variables $u$, $s$ and $w$ on "initial value configurations" of the lattice points, which through the eqs. (4.1a)-(4.1e) are updated in an unambiguous and consistent way when moving through the lattice. As an inspiration for the lattice case it is simpler to investigate first what happens in a semi-discrete limit of the lattice system in order to see clearer for what variables initial data have to be provided such that we have a welldefined and consistent evolution of the the data.

Through the continuum limit $\delta \to 0$ , $\delta = a - b$, employing the Taylor expansions:

$$\widehat{u} \;\to\; \widetilde{u}+\delta\dot{\widetilde{u}}+\ldots \quad , \quad \widehat{s} \;\to\; \widetilde{s}+\delta\dot{\widetilde{s}}+\ldots \quad , \quad \widehat{w} \;\to\; \widetilde{w}+\delta\dot{\widetilde{w}}+\ldots \quad ,$$

in which the dot $\dot{}$ denotes the derivative w.r.t. a continuous time-variable $t$, we obtain the differential-difference system:

$$(2a+\underset{\sim}{u}-\widetilde{u})(1+\dot{u})+f(\widetilde{s}-\underset{\sim}{s})\dot{s}=2a \tag{4.2a}$$

$$[(a+\underset{\sim}{u})\widetilde{s}+(a-\widetilde{u})\underset{\sim}{s}]\dot{s}=(\widetilde{s}-\underset{\sim}{s})(\dot{w}+s) \tag{4.2b}$$

$$[\widetilde{w}-\underset{\sim}{w}-(a-u)\underset{\sim}{s}-(a+u)\widetilde{s}]\dot{s}=(\underset{\sim}{s}-\widetilde{s})s(1+\dot{u}) \tag{4.2c}$$

$$\left(a+u-\frac{\widetilde{w}}{\widetilde{\widetilde{s}}}\right)\left(a-\widetilde{u}+\frac{w}{s}\right)=a^2-\left(\frac{1}{s\widetilde{s}}+3e+fs\widetilde{s}\right) \tag{4.2d}$$

$$\left(1+\frac{\dot{w}}{s}-\frac{w\dot{s}}{s^2}\right)\left(a-u+\frac{\underset{\sim}{w}}{\underset{\sim}{s}}\right)+\left(a+\underset{\sim}{u}-\frac{w}{s}\right)(1+\dot{u})=2a-\frac{\dot{s}}{s^2\underset{\sim}{s}}+f\underset{\sim}{s}\dot{s} \tag{4.2e}$$

It is easy to check that eq. (4.2e) is redundant as the substition of $\dot{u}$, $\dot{s}$, $\dot{w}$ from (4.2a), (4.2b) respectively (4.2c) leads to a triviality. Furthermore, taking the $t$-derivative of (4.2d) and back-substituting the $t$-derivatives from (4.2a)-(4.2c), making use also of (4.2d) as it stands, we get again a triviality. Thus, the D$\Delta$-system (4.2) represents a consistent time-evolutionary system, for which a well-posed initial value problem can be formulated by assigning at $t = 0$ all values of of $u$ and $s$ along the one-dimensional chain, and providing one value of $w = w_0$, say at $n = 0$ at $t = 0$. Essentially this means that we are dealing with a 2-component D$\Delta$ system in terms of $u$ and $s$, for which the value $w_0$ acts as a background value. This is not really surprising, since we are dealing with a two-parameter extension of the *potential* lattice KdV equation and not of the KdV itself. We expect that the dependence of the IVP on the background value $w_0$ will disappear if one formulates the IVP for the properly defined extension of the KdV lattice itself.

Let us now return to the fully discrete lattice system, where we can recognise a similar situation. Motivated by the KdV case, cf. [21], it is natural to investigate a *local* iteration scheme (i.e. one for which each iteration step at a given vertex involves initial data at only a finite configuration of points) is given on "staircases", as in Figure 2, assigning values for $u$ and $s$ on the vertices of this staircase, and considering the discrete-time shift to be the map $(u_i, s_i) \mapsto (\widehat{u}_i, \widehat{s}_i)$. From the consideration of the semidiscrete case we expect that we need in addition only one "background" value $w_0$ at a specific point on the staircase. Indeed, setting the IVP up in this way it can be shown by straightforward computation that it is well-posed. In fact, from the three first equations in (4.1), namely (4.1a), (4.1b) and (4.1c), one can solve $\widehat{\widetilde{u}}$, $\widehat{\widetilde{s}}$, and $\widehat{\widetilde{w}}$ in a unique way, given the values of the other variables $u$, $\widetilde{u}$, $\widehat{u}$ as well as $s$, $\widetilde{s}$, $\widehat{s}$ and $w$, $\widetilde{w}$, $\widehat{w}$. The equations (4.1d) and (4.1e) link the variables $\widetilde{w}$ and $\widehat{w}$ to $w$ and to the $u$,$s$-variables at the relevant lattice points. Thus, it remains to be verified that the shifted forms $\widehat{(4.1d)}$ and $\widetilde{(4.1e)}$ trivialise through back-substitution of $\widehat{\widetilde{w}}$ which was already obtained, and this is readily done. Furthermore, it is easily checked that the two ways of calculating $\widehat{\widetilde{w}}$ from either (4.1d) followed by $\widetilde{(4.1e)}$, or from (4.1e) followed by $\widehat{(4.1d)}$ are consistent.

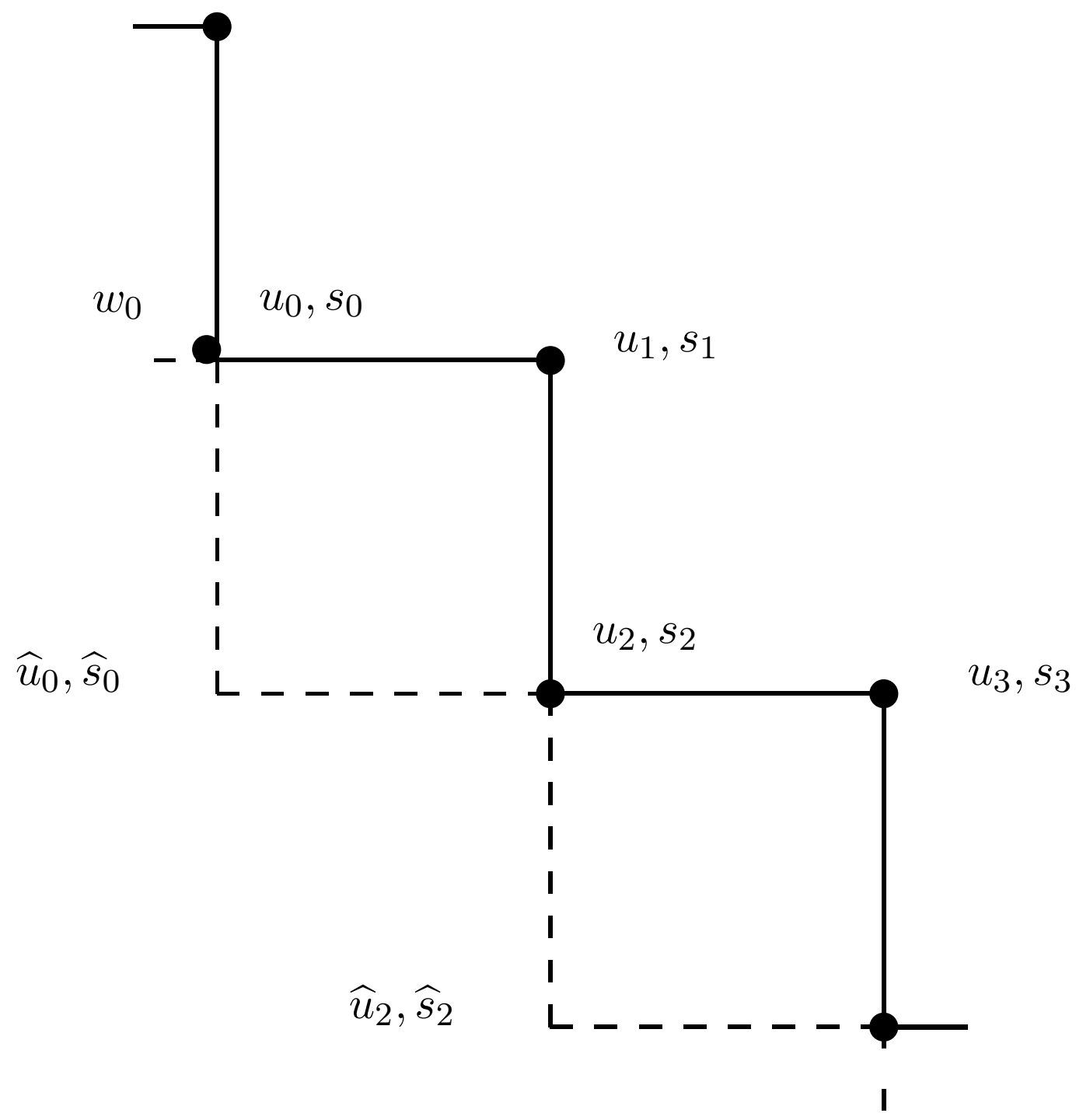

**Figure 2:** *Assignment of initial values on staircases in the lattice.*

We may conclude from these considerations that the coupled system (4.1) is in effect a system for for $u$ and $s$, with one of the eqs. (e.g. (4.1c) being redundant, whilst the eqs. for $w$ (eqs. (4.1d) and (4.1e)) are consistent with the system (4.1a)+(4.1b) involving effectively only a background value (or "forcing") through the variable $w_0$.

# 5 Associated Continuous Systems

As was demonstrated in the past for the lattice systems studied in [17, 23, 14], there exist many compatible continuous systems associated with them. These form, in fact, the continuous symmetries for the lattice systems (whilst the lattice systems constitute the discrete symmetries for the corresponding continuous flows). We will give here a few of the simplest of such associated continuous flows for the purpose of identification of the associated lattice system.

Starting from the continuous relations (2.14) we can derive now a set of partial diferential equations which are compatible with the discrete equations. Taking into account that the derivatives w.r.t. the *even* time-flows $x_{2j}$ ($j \in \mathbb{Z}$) all vanish by construction (as a consequence of the infinite matrix $\boldsymbol{U}$ being symmetric), we concentrate on the first two nontrivial time-flows in terms of $x := x_1$ and $t := x_3$. Thus, from the basic relations (2.14) for $j = 1, 3$, which can be cast into the forms:

$$\boldsymbol{U}_x = \boldsymbol{\Lambda}\cdot\boldsymbol{U} + \boldsymbol{U}\cdot {}^t\boldsymbol{\Lambda} - \boldsymbol{U}\cdot\left(\boldsymbol{O} - f\,{}^t\boldsymbol{P}^{-1}\cdot\boldsymbol{O}\cdot\boldsymbol{P}^{-1}\right)\cdot\boldsymbol{U} \tag{5.1a}$$

$$\boldsymbol{U}_t = \frac{1}{4}\boldsymbol{U}_{xxx} + \frac{3}{2}\boldsymbol{U}_x\cdot\left(\boldsymbol{O} - f\,{}^t\boldsymbol{P}^{-1}\cdot\boldsymbol{O}\cdot\boldsymbol{P}^{-1}\right)\cdot\boldsymbol{U}_x \tag{5.1b}$$

together with the relation (2.13) we find the following set of relations in terms of the objects defined in section 3:

$$2U_{0,1} = u_x + u^2 - fs^2 \quad , \quad fh - u = \frac{s_x + v - w}{s} \tag{5.2}$$

$$2U_{-1,-2} = h_x + s^2 - fh^2 \quad , \quad U_{1,-1} = \frac{1 - vw}{s} \tag{5.3}$$

$$-v_x = \frac{1 - vw}{s} + 3es + v(u + fh) + fs(h_x + s^2 - fh^2) \tag{5.4}$$

$$w_x = \frac{1 - vw}{s} + 3es + w(u + fh) - s(u_x + u^2 - fs^2) \tag{5.5}$$

for the $x$-derivatives, and after eliminating higher powers of $\boldsymbol{\Lambda}$ and ${}^t\boldsymbol{\Lambda}$ in the infinite-matrix relations for the $t$-derivatives we obtain:

$$u_t = \frac{1}{4}u_{xxx} + \frac{3}{2}u_x^2 - \frac{3}{2}fs_x^2 \tag{5.6a}$$

$$s_t = \frac{1}{4}s_{xxx} + \frac{3}{2}s_x u_x - \frac{3}{2}fh_x s_x \tag{5.6b}$$

$$h_t = \frac{1}{4}h_{xxx} + \frac{3}{2}s_x^2 - \frac{3}{2}fh_x^2 \tag{5.6c}$$

$$v_t = \frac{1}{4}v_{xxx} + \frac{3}{2}v_x u_x + \frac{3}{2}fs_x(U_{-1,-2})_x \tag{5.6d}$$

$$w_t = \frac{1}{4}w_{xxx} + \frac{3}{2}s_x(U_{0,1})_x - \frac{3}{2}fw_x h_x \ . \tag{5.6e}$$

In order to get the lowest-order system, we eliminate the dependent variables $h$ and $v$ and obtain a system solely in terms of $u$, $s$ and $w$, which can be simplified further by introducing the quantity $A = -u + w/s$ . Thus, we obtain finally the coupled systems of nonlinear evolution equations:

$$s_t = \frac{1}{4}s_{xxx} + \frac{3}{2}s_x\left[\frac{1}{s^2} + 3e + fs^2 - A^2 + A\frac{s_x}{s} - \frac{1}{2}\frac{s_{xx}}{s}\right] \tag{5.7a}$$

$$A_t = \frac{1}{4}A_{xxx} - \frac{3}{2}A^2A_x + \frac{3}{2}A_x\left(\frac{1}{s^2} + 3e + fs^2\right) + \frac{3}{4}\frac{s_x}{s}\left(\frac{1}{s^2} + 3e + fs^2\right)_x . \tag{5.7b}$$

Although by the introduction of the variable $A$ we have effectively eliminated $u$ and $w$, we mention the following relation among the original variables:

$$\left(u + \frac{w}{s}\right)_x + \left(u - \frac{w}{s}\right)^2 = P(s^2) = \frac{1}{s^2} + 3e + fs^2 \ , \tag{5.8}$$

which can be thought of as the continuous analogue of (1.2d). The coupled system (5.7) is a continuous analogue of the lattice system (1.2), which can be obtained from it by a (rather subtle) continuum limit. This system is integrable in its own right in the sense of admitting a continuous Lax pair of the form:

$$\phi_x = \begin{pmatrix} k - u + \frac{f}{K}sw & 1 - \frac{f}{K}s^2 \\ K + 3e - u_x - u^2 + fs^2 + \frac{f}{K}w^2 & k + u - \frac{f}{K}ws \end{pmatrix}\phi \tag{5.9a}$$

$$\begin{aligned} \phi_t &= k^2\phi_x - \begin{pmatrix} (u_{0,1})_x & u_x \\ (u_{1,1})_x & -(u_{0,1})_x \end{pmatrix}\phi + \\ &- \frac{f}{K}\begin{pmatrix} (1-vw)\frac{s_x}{s} + v_x w & vs_x - sv_x \\ (1-vw)\frac{w_x}{s} - w\left(\frac{1-vw}{s}\right)_x & (vw-1)\frac{s_x}{s} - v_x w \end{pmatrix}\phi \end{aligned} \tag{5.9b}$$

with substitutions:

$$u_{0,1} = \frac{1}{2}(u_x + u^2 - fs^2) \quad , \quad (u_{1,1})_x = \frac{1}{2}(u_{0,1})_{xx} + u(u_{0,1})_x - fsw_x \ .$$

We have not been able to trace the system of equations (5.7) in the literature, but we note the amusing fact that if we set $A = \frac{1}{2}\frac{s_x}{s}$, then the system reduces to

$$s_t = \frac{1}{4}s_{xxx} - \frac{3}{4}\frac{s_x s_{xx}}{s} + \frac{3}{8}\frac{s_x^3}{s^2} + \frac{3}{2}s_x\left(\frac{1}{s^2} + 3e + fs^2\right) \tag{5.10}$$

which is the famous second modified KdV (M$^2$KdV) equation, first found by Calogero and Degasperis, cf. [6], and also [12, 8]. However, we must point out that the coupled system (5.7) is richer than the M$^2$KdV equation, which can be verified immediately by reconsidering the soliton solutions (3.14) presented in section 3.2. In fact, taking the factors $r_i$ of (3.12) to include factors of the form $\exp(2k_ix + 2k_i^3t)$ , it is easily verified that the formulae (3.14) provide an explicit solution of (5.7) with $A = -u+w/s$. However, it is readily checked that this does *not* provide a solution of (5.10), nor does the relation $A = s_x/(2s)$ hold for these soliton solutions!

It is known that the third modified KdV (M$^3$KdV) equation, cf. [11] and also [22], is expressible in terms of elliptic functions, cf. also [24]. However, that equation seems to have quite different structure from (5.7), and its soliton solutions, [11], do not seem to involve an elliptic curve. Thus, we are tempted to believe that the coupled system (5.7) and its lattice counterpart (1.2) forms a genuine 2-parameter "elliptic" extension of the continuous and lattice potential KdV equations.

# 6 Discussion

In this paper we have presented a general scheme to obtain integrable systems (i.e. partial difference equations) associated with elliptic curves. These systems constitute a *2-parameter deformation* of the lattice systems that were investigated in the past. For the case $f = 0$, i.e. when the elliptic curve generates into a rational curve, we recover the usual lattice KdV system that was studied from various perspectives during the last decades. One relevant recent discovery in which these systems played an instrumental role, is that the KdV hierarchy –when formulated appropriately– is rich enough to contain the full Painlevé VI equation as similarity reduction. This slightly surprising observation was made in a recent paper [18] and subsequently it was shown in [16] that the entire KdV hierarchy can be encoded in one single nonautonomous PDE which is also reducible to the full PVI equation with arbitrary parameters. This demonstrates that these systems are in some sense universal. The programme that now unfolds on the basis of the results of the present paper, is to develop similar systems associated with an elliptic curve, which would then constitute 2-parameter deformations of the systems just mentioned. In that way one expects that these systems under appropriate similarity reductions reduce to an "ellipticised" version of PVI, possibly the equations that were derived in the 1970's by K. Okamoto from isomonodromic deformation problems on the torus, cf. [20].

We should mention that there exists also an alternative way to extend the lattice systems of KdV type such that there is an underlying elliptic curve. In fact, V. Adler discovered in [2] a lattice version of the famous Krichever-Novikov equation, cf. [10]. The resulting lattice system has the following form:

$$\begin{aligned} &k_0 u\widetilde{u}\widehat{u}\widehat{\widetilde{u}} - k_1\left(u\widetilde{u}\widehat{u} + u\widetilde{u}\widehat{\widetilde{u}} + u\widehat{u}\widehat{\widetilde{u}} + \widetilde{u}\widehat{u}\widehat{\widetilde{u}}\right) + k_2\left(\widetilde{u}\widehat{u} + u\widehat{\widetilde{u}}\right) \\ &-k_3\left(u\widetilde{u} + \widehat{u}\widehat{\widetilde{u}}\right) - k_4\left(u\widehat{u} + \widetilde{u}\widehat{\widetilde{u}}\right) + k_5\left(u + \widetilde{u} + \widehat{u} + \widehat{\widetilde{u}}\right) + k_6 = 0 \ , \end{aligned} \tag{6.1}$$

where the coefficients $k_0, \ldots, k_6$ are parametrised through elliptic functions. The main difference between (6.1) and (1.2) resides in the fact that the lattice parameters for the Adler equation are themselves points of the elliptic curve, namely with coordinates $(\wp(\alpha), \wp'(\alpha))$, $(\wp(\beta), \wp'(\beta))$, whereas this is not the case for the system (1.2). Furthermore, the Lax pair for (6.1), which was constructed in [13], has the spectral parameter living also on the elliptic curve, whereas this is not evident for the Lax pair for (1.2). Nonetheless, the explicit formulae for soliton solutions which were discussed in subsection 3.2, demonstrate very clearly that they contain parameters $(k_i, K_i)$ which live on the elliptic curve. Thus, this seems clear indication that the elliptic curve (2.6) is essential, and that the lattice system (1.2) constitues a genuine 2-parameter extension of the lattice (potential) KdV equation (1.1). A further comparison of between the two lattice systems is still needed as well as a clear identification of the underlying continuous system (5.7).